\documentstyle[12pt,a4,psfig,epsf]{article}
\newcommand{\bqa}{\begin{eqnarray}}
\newcommand{\eqa}{\end{eqnarray}}
\newcommand{\beq}{\begin{eqnarray}}
\newcommand{\eeq}{\end{eqnarray}}

\begin{document}
\vskip 20mm
\centerline{\Large\bf {Bose-Einstein Condensation in}}
\centerline{\Large \bf {Anisotropic Harmonic Traps}}  
\vskip 10mm
\centerline{T. Haugset, H. Haugerud and J. O. Andersen}
\centerline{{\it Institute of Physics}}
\centerline{{\it University of Oslo}}
\centerline{{\it N-0316 Oslo, Norway}}

\vskip 10mm
\begin{abstract}
We study the thermodynamic behaviour of 
an ideal gas of bosons trapped in a three-dimensional anisotropic harmonic 
oscillator potential.
The condensate fraction as well as the specific heat is calculated
using the Euler-Maclaurin approximation. For a finite number of 
particles there is no phase transition, but
there is a well defined temperature at which
the condensation starts. 
We also consider condensation in lower dimensions, and for one-dimensional
systems we discuss the dependence of the condensate fraction and heat capacity
on the ensemble used.
\end{abstract}
\normalsize

\section{Introduction}
The observation of Bose Einstein condensation (BEC)
in ultracold alkali gases has 
led to intensive experimental and theoretical investigations of 
the phenomenon \cite{WWW}. 
Since the atomic gases of the experiments are dilute, and hence weakly 
interacting, 
their physical properties may be explored theoretically in terms of basic
interatomic interactions and as a first approximation the interactions may  
be neglected~\cite{sod}. 
This is strongly opposed to the physical properties of related  
phenomena such as 
superconductivity in metals and superfluidity in liquid He, which are
fundamentally modified by strong interparticle interactions.

In the remarkable pioneering experiment \cite{AEMWC} ${}^{87}$Rb vapour was 
magnetically trapped and evaporatively cooled to approximately
$170$ nK, where a finite 
fraction of the 
particles was observed to occupy the ground state. When 
further lowering the temperature this fraction increased abruptly, 
signalling a Bose-Einstein condensation. 
The confining magnetic field was a 
time-averaged, orbiting potential \cite{PETRICH} whose effective potential 
to first order is a three-dimensional anisotropic harmonic oscillator. The 
potential was not anisotropic in all three directions, but was cylindrically 
symmetric.
Shortly after, a similar experiment was performed with ${}^{23}$Na gas \cite{sod}; 
in 
this case the potential was completely anisotropic.

With these experiments in 
mind, 
we will in the present
paper extend our previous work on the thermodynamics of an 
ideal 
isotropic gas \cite{HR,HHR} to the anisotropic case.


Although the interactions among the atoms are weak, they are quite important 
in the 
condensate. Both experiments and theory show that due to the interaction the 
actual ground state is considerably larger 
than the single particle ground state.
However, close to the transition temperature just a few particles 
occupy the 
ground state and the gas is much more dilute. At these temperatures one would 
believe 
that the ideal gas is a good approximation.

In Ref.~\cite{KvD} it has been proposed that highly anisotropic traps may
freeze out one or two dimensions, so that the system to a good approximation
is lower dimensional. This implies that it is of interest to study 
these systems also, and we present a preliminary analysis of them
in this work.

The plan of the article is as follows. In the next section we study 
noninteracting
bosons in a confining anisotropic harmonic potential in three dimensions 
(3$D$).
In the 
framework of the grand canonical ensemble we calculate the condensate fraction, 
the internal energy and the heat capacity analytically using Euler-Maclaurin summation 
and compare with exact numerical summation.
In section three we discuss the same system in one and two spatial dimensions.
In particular, we study the effect of 
using different ensembles in one-dimensional systems.
In section four we summarize and conclude.

\section{Bose-Einstein Condensation in Three Dimensions}
In this section we will discuss Bose-Einstein condensation in three-dimensional
harmonic traps.
Firstly, we consider the isotropic potential and then we move on to the
anisotropic trap. However, we would first like to discuss Bose-Einstein
condensation and phase transitions in a more general setting, and comment
upon earlier work.

Let us begin our discussion by considering free bosons in 3$d$ in the 
thermodynamic limit ($N,V\rightarrow \infty$, $N/V=\rho$ finite).
In this case one may show (see any standard textbook in statistical mechanics)
that the fraction of particles in the ground state
is given by 
\begin{eqnarray}
\frac{N_{0}}{N}=1-\Big(\frac{T}{T_{c}}\Big)^{3/2},
\end{eqnarray}
where $T_{c}$ is defined through
\begin{eqnarray}
\rho\left(\frac{2\pi\hbar^{2}}{mk_BT_{c}}\right)^{3/2} = \zeta(3/2).
\end{eqnarray}
As can be seen from the above equation, $T_{c}$ defines the onset of 
the condensation of particles into the ground state.
Moreover, it can be demonstrated that the chemical potential vanishes for
$T\leq T_{c}$ and that $\mu <0$ for $T>T_{c}$. Thus, one can {\it define}
the critical temperature to be the highest temperature for which $\mu =0$.
The derivative of $\mu$ with respect to temperature is discontinuous at
$T=T_{c}$, and so there is a first order phase transition.\\ \\
Generally, the chemical potential must
satisfy $\mu\leq E_{0}$, where $E_{0}$ is the ground state energy of the
system, in order to have positive occupation numbers. In~\cite{KT3}, 
Kirsten and Toms propose that a necessary criterion for Bose-Einstein 
condensation is that $\mu$ must reach
the ground state energy $E_{0}$ at finite temperature.
From this, they show that BEC only occurs for
free bosons when $D\geq 3$,\footnote{Or more precisely that $T_{c}=0$ when 
$D\leq 2$.} 
and not for free bosons
in a finite volume in three dimensions.
In a system with a constant magnetic field
it only happens in five or more dimensions. This latter result has also
been verified independently by Elmfors {\it et al}~\cite{ELM1}.\\ \\
From theoretical work as well as from an experimental point of view 
we will argue that this criterion is somewhat too strict. 
In Ref.~\cite{HHR} it has been demonstrated numerically that the chemical
potential never reaches its critical value in the case of bosons in a 
confining harmonic potential. According to the criterion mentioned above,
this system does not exhibit BEC.
However, the occupation number of the ground
state changes considerably in a narrow temperature range, and attains
a macroscopic value. More specifically, it has been shown 
(see e.g.~\cite{HHR}) that the 
condensate near 
$T_{0}=\frac{\hbar\omega}{k_{B}}\Big(\frac{N}{\zeta (3)}\Big)^{1/3}$
varies as 
\begin{eqnarray}
\frac{N_{0}}{N}=1-\Big(\frac{T}{T_{0}}\Big)^{3}+O(N^{-1/3}).
\end{eqnarray}
This abrupt change in number of particles in the ground state takes place
near the temperature at which the 
heat capacity has a pronounced maximum (which is $T_{0}$). 
We would not hesitate to call
this a condensation, although there is no critical behaviour and so it
is not obvious how to define a critical temperature. However, in Ref.~\cite{path} it has been proposed that the critical temperature is defined by the maximum of the specific heat. According to this criterion there is a phase transition in two and three dimensions, but not in 
one (see also section three).
We think is it natural to call the temperature $T_{0}$
the {\it condensation temperature}, since this temperature determines the
onset of condensation of particles into the ground state.
This view is supported by experiment, namely the fact that one has measured
a large increase in the number of
particles in the ground state over a narrow temperature region.

\subsection{Isotropic Potential}
The eigenfunctions of the three-dimensional harmonic oscillator 
Hamiltonian can  always
be written as products of three eigenfunctions of the one-dimensional 
oscillator, which are Hermite polynomials.
Hence, the eigenfunctions can be labeled by three integers $n_{x}$, $n_{y}$
and $n_{z}$ and the energy is $\hbar\omega (n_{x}+n_{y}+n_{z})$ in the
isotropic case (we have set the ground state energy to zero).
Each energy level can thus be characterized by a single
quantum number $n$ and the corresponding
degeneracy is $g_n = (n+1)(n+2)/2$, which simply is
the number of ways of writing $n$ as the sum of the three positive integers
$n_{x}$, $n_{y}$ and $n_{z}$. 
The quantum number $n$ then takes the values $n=0,1,2,\ldots$. 
Due to the high degree of symmetry of the isotropic harmonic well, 
there is another convenient basis set. The Hamiltonian commutes with
$\hat{L}^{2}$ and $\hat{L}_{z}$ and hence the eigenfunctions can be 
labeled by $L$,
$L_{z}$ and the energy. The eigenfunctions are Laguerre polynomials
times a spherical harmonic.\\ \\
At temperature $T$ and chemical potential $\mu$ the particle number 
$N$ is given by 
\begin{eqnarray}
N = \sum_{n=0}^{\infty}\frac{g_n}{e^{\beta(\varepsilon_n - \mu)} - 1},
                                                             \label{free:1}
\end{eqnarray}
where $\beta=1/k_BT$ and $g_n/(e^{\beta(\varepsilon_n - \mu)} - 1)$ is the
usual Bose-Einstein distribution function.
Introducing the fugacity $\lambda = \exp(\beta\mu)$, the ground state 
particle number may be written as
\begin{eqnarray}
N_0 = \frac{\lambda}{1 - \lambda}.
\end{eqnarray}
The number of particles in excited states is then
\begin{eqnarray}
N_e = \sum_{n=1}^{\infty}\frac{g_n\lambda e^{-bn}}{1 - \lambda e^{-bn}},
\end{eqnarray}
which is a function of the {\em effective fugacity} $\bar{\lambda} 
= \lambda e^{-b}$ where $b = \beta\hbar\omega$.\\

Since the particle number 
is given, one 
must express the chemical potential as a function of temperature and 
particle number by inverting Eq. (\ref{free:1}) 
in order to derive thermodynamic quantities describing the system.
Of particular interest to
us is the particle number in the ground state as well as the specific heat 
of the system as a function of temperature.

Now let us move on to actual calculations and consider different
ways of handling Eq.~(\ref{free:1}).
At high temperature $b\ll 1$, the difference between the terms in the 
sum is small, and one can approximate the sum by an integral. 
The dominating contribution at high $T$ is found by setting 
$g_n \simeq n^2/2$, leading to the semiclassical limit \cite{HR,BPK}. 
The transition temperature, as defined by the temperature at 
which $\lambda = 1$, is in this approximation, found to be \cite{HR}
\begin{eqnarray}
T_0 = \frac{\hbar\omega}{k_B}\left(\frac{N}{\zeta(3)}\right)^{1/3}.
\end{eqnarray}
In the same approximation the specific heat $C_V$ exhibits a 
discontinuity at this temperature with the shape of a $\lambda$. 
The discontinuity is however an artifact of the approximation. 
An exact numerical calculation \cite{HHR} reveals that $C_V$ is a 
continuous function of $T$. This is a consequence of the finiteness 
of the particle number. 

As discussed in \cite{HHR}, an efficient way of performing summations 
is given by the Euler-Maclaurin summation method \cite{PFTV}. The sum 
is then written as an integral plus an infinite series of correction terms
\begin{eqnarray}
\sum_{n=a}^{n=b}f(n) = \int_a^bdxf(x) + \frac{1}{2}[f(b) + f(a)] 
+ \frac{1}{12}[f'(b) - f'(a)] + \cdots
\end{eqnarray}
Applying this formula to the sum over excited states, 
the contributions 
from the upper limit vanish
(it is natural to single out the ground state from the sum
and use Euler-Maclaurin on the rest. Moreover, the series converges much 
faster doing so). Higher order derivatives mainly bring down 
higher powers of $\hbar\omega/k_BT$, although differentiation of the 
degeneracy factor yields small corrections to lower order. At high 
temperature we may therefore safely truncate the series. Including 
only the first derivative, the number of excited particles becomes
\begin{eqnarray}
N_e = G_3(\bar{\lambda}) + \frac{3}{2}G_2(\bar{\lambda}) + G_1(\bar{\lambda}) + 
\frac{1}{4}\frac{\bar{\lambda}}{1 - \bar{\lambda}}\left(\frac{31}{6} + \frac{b}{1 - \bar{\lambda}}\right).
\end{eqnarray}
We have here introduced the functions
\begin{eqnarray}
\label{lower}
G_{p+1}(\bar{\lambda}) = \frac{1}{p!}\int_1^{\infty}dx\frac{x^p\lambda 
e^{-bx}}{1 - \lambda e^{-bx}}.
\end{eqnarray}
These functions satisfy a simple recursion relation
\begin{eqnarray}
G_{p+1}(\bar{\lambda}) = -\frac{1}{bp!}\ln(1 - \bar{\lambda}) + \frac{1}{b}\int_0^{\bar{\lambda}}
\frac{d\lambda}{\lambda}G_p(\lambda).
\end{eqnarray}
They can be expressed in terms of the more familiar polylogarithmic functions 
\begin{eqnarray}
\label{polylog}
\mbox{Li}_p(z) = \sum_{n=1}^{\infty}\frac{z^n}{n^p},
\end{eqnarray}
which are generalizations of the Riemann $\zeta$-function. For $p=0, 1,$ 
and $2$ we have
\begin{eqnarray}\nonumber
G_{1}(\bar{\lambda} )&=&\frac{1}{b}\mbox{Li}_{1}(\bar{\lambda} )\\ \nonumber
G_{2}(\bar{\lambda} )&=&\frac{1}{b}\mbox{Li}_{1}(\bar{\lambda} )
+\frac{1}{b^2}\mbox{Li}_{2}(\bar{\lambda} )\\ 
G_{3}(\bar{\lambda} )&=&\frac{1}{2b}\mbox{Li}_{1}(\bar{\lambda} )
+\frac{1}{b^2}\mbox{Li}_{2}(\bar{\lambda} )
+\frac{1}{b^3}\mbox{Li}_{3}(\bar{\lambda} ).
\end{eqnarray}
The $\mbox{Li}$-functions are obtained when the lower limit of the integral in 
Eq.~(\ref{lower})
is 
set to zero. For $n\leq 1$ they may be written in terms of more 
familiar functions, e.g. $\mbox{Li}_1(z) = -\ln(1 - z), 
\mbox{Li}_0(z)  = z/(1 - z)$. The polylogarithms satisfy the functional relation
\begin{eqnarray}
\frac{d\mbox{Li}_p(z)}{dz} = \frac{1}{z}\mbox{Li}_{p-1}(z).
\end{eqnarray}
In terms of these functions, the total number of particles reads
\begin{eqnarray}
N_e &=& \frac{\lambda}{1 - \lambda} + \frac{1}{b^3}\mbox{Li}_1(\bar{\lambda}) 
+ \frac{5}{2b^2}\mbox{Li}_2(\bar{\lambda}) + \frac{3}{b}\mbox{Li}_1(\bar{\lambda})\nonumber\\
&+& \frac{1}{4}\frac{\bar{\lambda}}{1 - \bar{\lambda}}\left(\frac{31}{6} 
+ \frac{b}{1 - \bar{\lambda}}\right).
\end{eqnarray}
The internal energy as well as the heat capacity have been derived in a similar
way in Ref.~\cite{HHR}.
\subsection{Anisotropic Potential}
The potentials used in the experiments~\cite{sod,AEMWC} 
are either partially or 
completely asymmetric. We therefore need expressions for the cases 
where the frequencies are different. 
Let us first consider the case $\omega_x=\omega_y\neq\omega_z$.
As for the isotropic potential we use the 
Euler-Maclaurin formula for each of the sums. The calculations are similar
and we skip the details.
The result is
\begin{eqnarray}
N &=& \frac{\lambda}{1 - \lambda} + \frac{1}{b_x^2b_z}\mbox{Li}_3(\lambda_{xz}) 
+ 
\left(\frac{1}{2b_x^2} + \frac{2}{b_xb_z}\right)\mbox{Li}_2(\lambda_{xz}) 
\nonumber\\
&&+ \left(\frac{b_z}{12b_x^2} + \frac{1}{b_x} + \frac{11}{12b_z}\right)
\mbox{Li}_1(\lambda_{xz})
+ \left(\frac{b_z}{6b_x} + \frac{11}{24} + \frac{b_x}{6b_z}\right)
\mbox{Li}_0(\lambda_{xz}) \nonumber\\
&&+\frac{1}{b_x^2}\mbox{Li}_2(\lambda_x) + 
\frac{2}{b_x}\mbox{Li}_1(\lambda_x) + \frac{11}{12}\mbox{Li}_0(\lambda_x)
\nonumber\\
&&+ \frac{1}{b_z}\mbox{Li}_1(\lambda_z) + \frac{1}{2}\mbox{Li}_0(\lambda_z) 
+ {\cal O}
(\mbox{Li}_{-1}).
\end{eqnarray}
We have here introduced the notation $\lambda_{xz} = 
\lambda\exp(-b_x-b_z)$ etc. where $b_{x}=\hbar\beta\omega_{x}$ and so on. 
We also need the internal energy and the result using the Euler-Maclaurin 
formula is
\begin{eqnarray}
U/T &=& \frac{3}{b_x^2b_z}\mbox{Li}_4(\lambda_{xz}) 
+ \left(\frac{5}{b_xb_z}  +\frac{2}{b_x^2}\right)
\mbox{Li}_3(\lambda_{xz})\nonumber\\
&+& \left(\frac{7b_z}{12b_x^2} + \frac{7}{2b_x} + \frac{35}{12b_z} \right)
\mbox{Li}_2(\lambda_{xz}) + \left(\frac{b_z^2}{12b_x^2} + \frac{5b_z}{12b_x} 
+ \frac{23}{12} + \frac{17b_x}{12b_z}\right)\mbox{Li}_1(\lambda_{xz})\nonumber\\
&+& \left(\frac{b_z^2}{6b_x} + \frac{79b_z}{144} + \frac{19b_x}{24} 
- \frac{b_z^3}{360b_x^2} + \frac{b_x^2}{12b_z}\right)\mbox{Li}_0(\lambda_{xz})
\nonumber\\
&+& \frac{2}{b_x^2}\mbox{Li}_3(\lambda_x) + \frac{3}{b_x^2}\mbox{Li}_2(\lambda_x) 
+ 2\mbox{Li}_1(\lambda_x) + \frac{5b_x}{4}\mbox{Li}_{0}(\lambda_x)\nonumber\\
&+& \frac{1}{b_z}\mbox{Li}_2(\lambda_z) + \mbox{Li}_1(\lambda_z) + \frac{5b_z}{12}
\mbox{Li}_0(\lambda_z) + {\cal O}(\mbox{Li}_{-1}).  
\end{eqnarray}

The specific heat is now found by differentiating $U$ with respect to $T$. 
In doing so we need $d\mu/dT-\mu/T$ which is found by differentiation 
of $N$.\\ \\
In the ${}^{23}$Na experiment all three frequencies were different~\cite{sod}. 
Thus, we need the expression for the particle number in this case also.
By applying the Euler-Maclaurin approximation one finds that
$N$ reads
\begin{eqnarray}
N &=& \frac{\lambda}{1 - \lambda} + \frac{1}{b_xb_yb_z}\mbox{Li}_3
(\lambda_{xyz}) + \frac{1}{2}\left(\frac{1}{b_xb_y} + \mbox{sym.}\right)
\mbox{Li}_2(\lambda_{xyz})\nonumber\\
&&+ \frac{1}{4}\left[\frac{1}{b_x} + \mbox{sym.} + \frac{1}{3}
\left(\frac{b_x}{b_yb_z} + \mbox{sym.}\right)\right]\mbox{Li}_1(\lambda_{xyz})
\nonumber\\
&&+ \frac{1}{8}\left[1 + \frac{1}{3}\left(\frac{b_x}{b_y} + \mbox{sym.}\right)
\right]\mbox{Li}_0(\lambda_{xyz})
+ \frac{1}{b_xb_y}\mbox{Li}_2(\lambda_{xy}) \nonumber\\
&&+ \frac{1}{2}\left(\frac{1}{b_x} 
+ \frac{1}{b_y}\right)\mbox{Li}_1(\lambda_{xy}) + \frac{1}{4}
\left[1 + \frac{1}{3}\left(\frac{b_x}{b_y} + \frac{b_y}{b_x}\right)\right]
\mbox{Li}_0(\lambda_{xy}) + \mbox{sym.}\nonumber\\
&&+ \frac{1}{b_x}\mbox{Li}_1(\lambda_x) + \frac{1}{2}\mbox{Li}_0(\lambda_x) + \mbox{sym.} 
+ {\cal O}(\mbox{Li}_{-1}),
\end{eqnarray}
Here, ``sym.'' means symmetrization, e.g.
\begin{eqnarray}
\frac{1}{b_{x}b_{y}}+\mbox{sym.}=\frac{1}{b_{x}b_{y}}+\frac{1}{b_{x}b_{z}}
+\frac{1}{b_{y}b_{z}}.
\end{eqnarray}
The two leading terms at high temperature are
\begin{eqnarray}
N = \frac{1}{b_xb_yb_z}\mbox{Li}_3(\lambda) + \frac{1}{2}\left(\frac{1}{b_xb_y} 
+ \frac{1}{b_xb_z} + \frac{1}{b_yb_z}\right)\mbox{Li}_2(\lambda).
                                                         \label{free:xx}
\end{eqnarray}
In \cite{GH} Grossmann and Holthaus considered a potential where 
$\omega_y = \sqrt{2}\omega_x$ and $\omega_z = \sqrt{3}\omega_x$. They constructed a continuous density of states based upon the two leading terms in the degeneracy. For the $\mbox{Li}_2$-term they found a coefficient $\gamma(k_BT/\hbar\omega)$ where $\omega = (\omega_x\omega_y\omega_z)^{1/3}$. A numerical summation gave $\gamma \simeq 1.6$. This numerical factor is now easily found from Eq. (\ref{free:xx}):
\begin{eqnarray}
\gamma = \left(\frac{3}{4}\right)^{1/3}\left(\frac{1}{\sqrt{2}} 
+ \frac{1}{\sqrt{3}} + \frac{1}{\sqrt{6}}\right) \simeq 1.538,
\end{eqnarray}
which agrees quite well with their approximate result. The same value 
for $\gamma$ has been obtained independently in \cite{KT2}. \\ \\
For the internal energy one finds
\begin{eqnarray}
U/T &=& \frac{3}{b_xb_yb_z}\mbox{Li}_4(\lambda_{xyz}) + 2\left(\frac{1}{b_xb_y} 
+ \mbox{sym.}\right)\mbox{Li}_3(\lambda_{xyz})\nonumber\\
&& + \frac{1}{4}\left[5\left(\frac{1}{b_x} + \mbox{sym.}\right) 
+ \frac{7}{3}\left(\frac{b_x}{b_yb_z} + \mbox{sym.}\right)\right]
\mbox{Li}_2(\lambda_{xyz})\nonumber\\
&& + \left[\frac{3}{4} + \frac{b_x}{3b_y} + \frac{b_x^2}{12b_yb_z} 
+ \mbox{sym.}\right]\mbox{Li}_1(\lambda_{xyz})\nonumber\\
&& + \frac{2}{b_xb_y}\mbox{Li}_3(\lambda_{xy}) + \frac{3}{2}\left(\frac{1}{b_x} 
+ \frac{1}{b_y}\right)\mbox{Li}_2(\lambda_{xy})  \nonumber\\
&& + \left(1 + \frac{b_x}{2b_y} 
+ \frac{b_y}{2b_x}\right)\mbox{Li}_1(\lambda_{xy}) + \mbox{sym.}
+ \frac{1}{b_x}\mbox{Li}_2(\lambda_x) \nonumber\\
&& + \mbox{Li}_1(\lambda_x) + \mbox{sym.} 
+ {\cal O}(\mbox{Li}_0).
\end{eqnarray}
Again the specific heat can be found by differentiation of this expression.

\subsection{Results}
With the formulas derived above at hand, we can make predictions for the 
actual physical systems within the ideal gas approximation. In doing so 
we have used the values for the oscillator frequencies and particle numbers 
as given in~\cite{sod,AEMWC}.

For the completely anisotropic system (${}^{23}$Na) the frequencies 
are 410, 235 and 745 Hz, and the particle number $N$ is 700 000. 
The frequencies for the ${}^{87}$Rb experiment are $\omega_{z}=240\pi$ Hz 
and $\omega_{x,y}=\omega_{z}/\sqrt{8}$, and the
particle number is 20 000. For these systems we have calculated both the 
condensate fraction and the specific heat using the Euler-Maclaurin 
approximation. The results are compared with those of direct numerical 
summation and the semiclassical approximation. The temperature is 
measured in units of the semiclassical transition temperature $T_0$. 
It is 1666 nK for the sodium gas and 127 nK for the rubidium gas.\\

In Fig. 1 we plot the condensate fraction as function of temperature. 
The numerical and analytical results agree well. For $N =$ 700 000 the transition is quite sharp, and the condensate nearly vanishes 
about 1\% below $T_0$. For the smaller particle number, the transition is
smoother.

\begin{figure}[tbh]
\begin{center}
\mbox{\psfig{figure=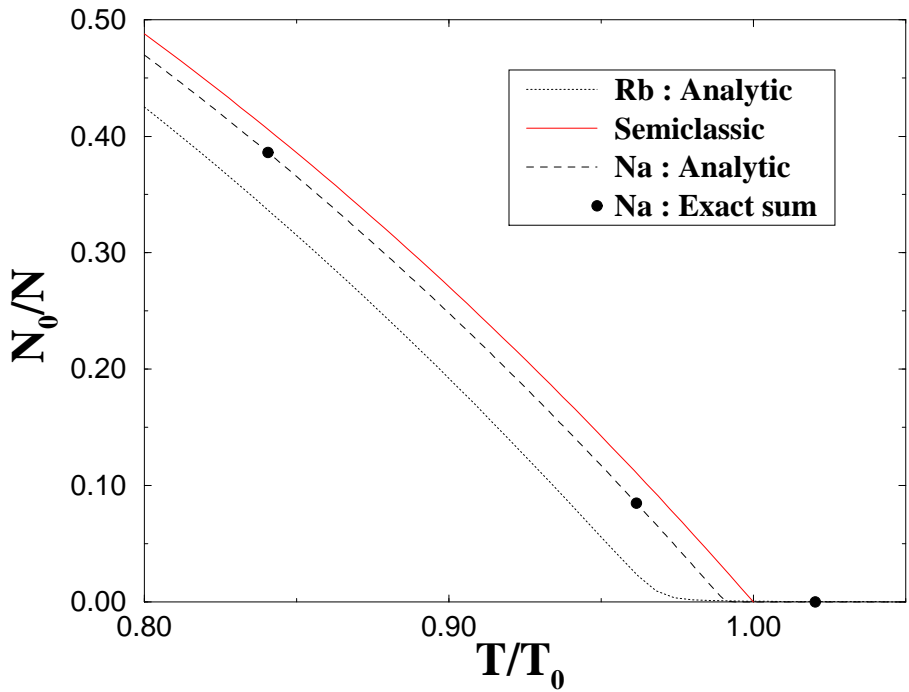,height=9cm,angle=270}}
\end{center}
\caption{\protect\footnotesize The condensate fraction $N_{0}/N$ in three dimensions 
with anisotropy-parameters taken from the ${}^{23}$Na and ${}^{87}$Rb
experiments. $N_{0}/N$ is plotted as a function of $T/T_{0}$.}
\label{3d1}
\end{figure}

The specific heat 
is shown in Fig. 2. We again find good agreement 
between numerical and analytical results. The discontinuity in the 
semiclassical approximation is smoothened by the corrections. 
Still, for $N =$ 700 000, the specific heat falls rapidly over a
temperature range of less than 10 nK. For this case it has a maximum 
value of $C_V/Nk_B \simeq 10.43$ , quite close to the 
semiclassical value $12\zeta(4)/\zeta(3)\simeq 10.80$. 
\begin{figure}[tbh] 
\begin{center}
\mbox{\psfig{figure=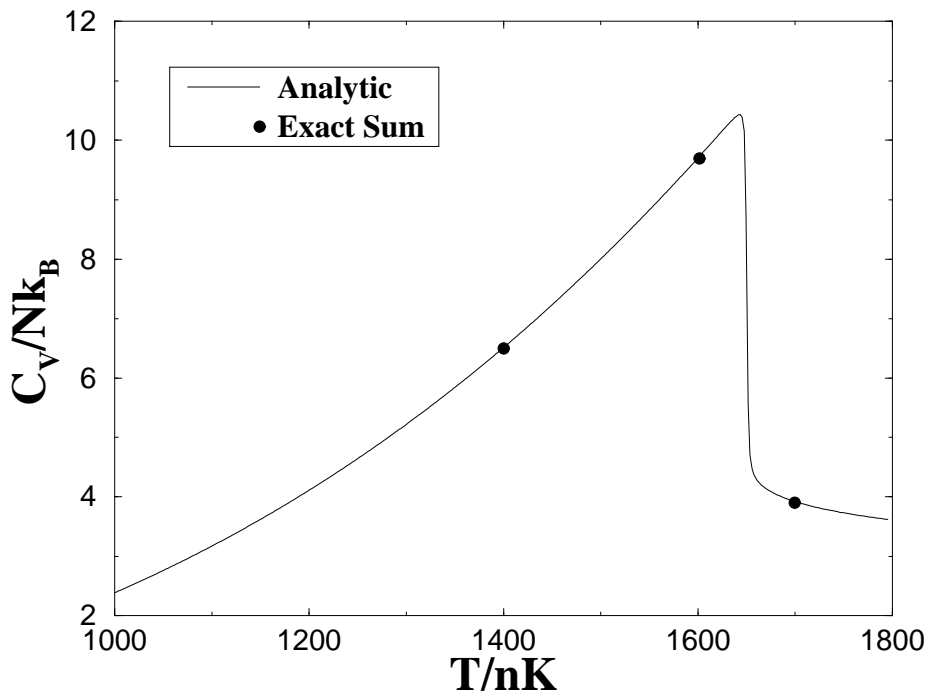,height=9cm,angle=270}}
\end{center}
\caption{\protect\footnotesize The specific heat $C_{V}$ in an ansiotropic three-dimensional 
trap with frequencies taken from the ${}^{23}$Na experiment. $C_{V}$ is plotted
as a function of $T/nK$.}
\label{3d2}
\end{figure}
\section{BEC in Lower Dimensions}
All BEC-experiments on trapped Bose gases reported so far are performed on 
three-dimensional systems. However, by the use of optical dipole traps or 
highly anisotropic magnetic traps one might, effectively, freeze out 
oscillations in one or two of the dimensions at low temperatures. 
In Ref.~\cite{KvD} the authors suggest the use of a field configuration 
with a radial energy level spacing of 200 nK, and a much smaller
axial spacing. At temperatures below 100 nK this system should 
effectively 
be one-dimensional.

In the previous section we discussed in some detail different criteria
for BEC. The same arguments apply here and we make a few comments.
For a free gas in the thermodynamic limit, the chemical potential vanishes
first at $T=0$. This implies that the critical temperature in one and two
dimensions is zero and that the condensate vanishes at finite temperature. 
In a finite volume there is again no critical behaviour and 
there is a condensate even at finite temperature.
This is also the case for an ideal gas trapped in an external harmonic
potential, as we 
shall see below.

In this section we consider the trapped Bose gas in both one and two 
dimensions using the same methods as for three dimensions. 
We start by examining the case of a highly anisotropic three-dimensional 
system, and demonstrate the effective reduction to one dimension 
when two of the oscillator frequencies are much higher than the third.
After this, we briefly discuss the use of different ensembles in one dimension.
As opposed to two and three dimensions it is here quite straightforward to
use the canonical ensemble.
For completeness, we also give some results for the two-dimensional case.

\subsection{Reduction to One Dimension}
Excitations in a given direction of the trap will be suppressed when the 
temperature is below the energy scale set by the corresponding oscillator 
strength. The highly anisotropic trap with $\omega_{x,y} \gg \omega_z$ should 
therefore effectively be one-dimensional at temperatures $k_{B}T \ll\hbar \omega_{x,y}$.
In Fig. 3 we have plotted the condensate fraction and specific heat $C_V$ 
for the case $\omega_{x,y} = 800 \omega_z$. In the same figure we also give 
the results for a one-dimensional system with frequency $\omega_z$.\\
\begin{figure}[tbh] 
\begin{center}
\mbox{\psfig{figure=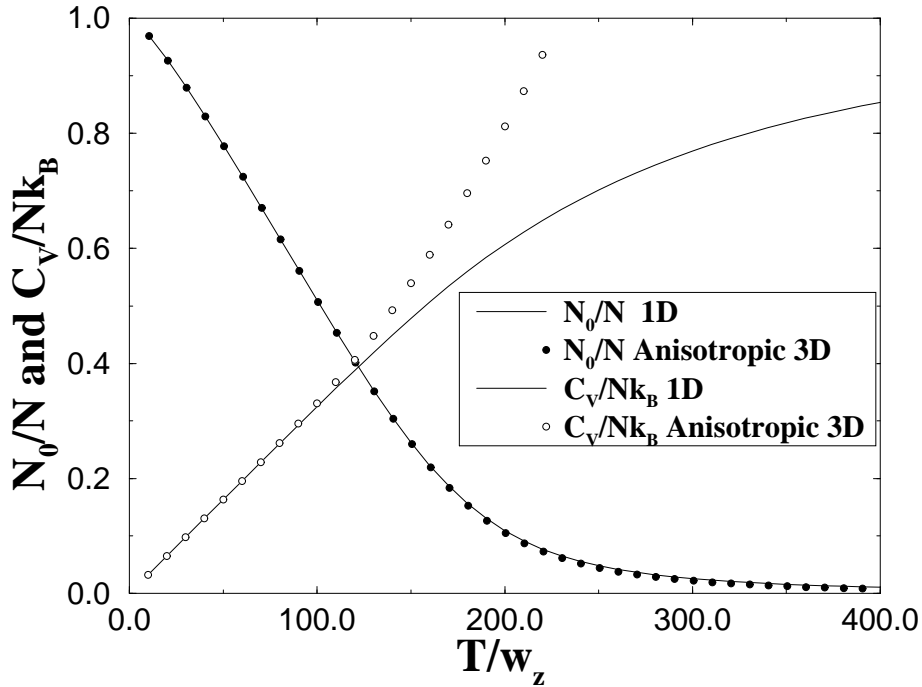,height=9cm,angle=270}}
\end{center}
\caption{\protect\footnotesize Comparison of the condensate fraction and the specific heat
of the highly anisotropic
trap and the one-dimensional trap.}
\label{3d1d}
\end{figure}

The three-dimensional condensate fraction is seen to agree very well with 
the one-dimensional one at low $T$, but at higher temperatures the former
is somewhat reduced due to radial excitations. 
These excitations have a more dramatic effect on the specific heat. The 
two curves have high temperature limits of three and one, respectively.

The expressions for the particle number and the energy density in
the one-dimensional case are given by:
\begin{eqnarray}\label{our}
N = \sum_{n=0}^{\infty}\frac{1}{e^{\beta(\varepsilon_n - \mu)} - 1} = N_0 
+ \frac{1}{b}\mbox{Li}_1(\bar{\lambda}) + \frac{1}{2}\mbox{Li}_0(\bar{\lambda}) + 
\frac{b}{12}\mbox{Li}_{-1}(\bar{\lambda}),
                                                        \label{eq:onedimN}
\end{eqnarray}
\begin{eqnarray}
U = \sum_{n=0}^{\infty}\frac{\varepsilon_n}{e^{\beta(\varepsilon_n - \mu)} - 1} 
= \omega\left[\frac{1}{b^2} \mbox{Li}_2(\bar{\lambda}) + \frac{1}{b}\mbox{Li}_1(\bar{\lambda}) 
+ \frac{5}{12}\mbox{Li}_0(\bar{\lambda}) + \frac{b}{12}\mbox{Li}_{-1}(\bar{\lambda})\right].
\end{eqnarray}

This one-dimensional system has earlier been studied by Ketterle and van 
Druten \cite{KvD}. For $\beta\omega\ll 1$, they obtained the equation
\begin{eqnarray}
\label{nn0}
N = N_0 + \frac{1}{b}\mbox{Li}_0(\lambda e^{-\beta\omega /2}).
\end{eqnarray}
The contribution from the excited states is similar to the first term in 
Eq.~(\ref{our}), but has a different effective fugacity. 
From Eq.~(\ref{nn0}), Ketterle and van Druten have defined a 
transition temperature
$T_{0}$ by setting $N_{0}=0$ and $\lambda =1$. We then have 
$N\approx \frac{k_{B}T_{0}}{\hbar\omega}\ln\frac{2k_{B}T_{0}}{\hbar\omega}$.
In Fig. 4 we compare 
various approximations for the density, plotting the condensate fraction as 
function of temperature for $N = 2\cdot 10^{4}$. Including all three terms in 
Eq. (\ref{eq:onedimN}), the analytical and numerical results agree to three 
decimal places. 
The heat capacity is a smooth function of temperature in the 
limit $N\rightarrow\infty$, and so there is no 
phase transition in one dimension.
\begin{figure}[tbh] 
\begin{center}
\mbox{\psfig{figure=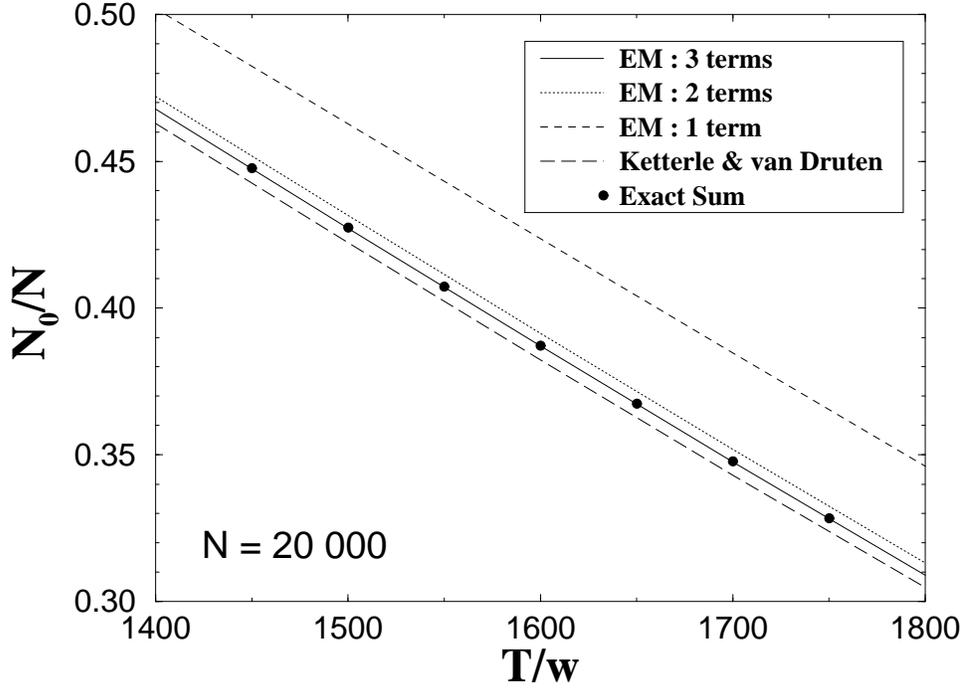,height=9cm,angle=270}}
\end{center}
\caption{\protect\footnotesize The condensate fraction in different approximations as a function of
$T/\omega$. $N=2\cdot 10^4$.}
\label{1d1}
\end{figure}

\subsection{Comparison of Ensembles in One Dimension}
The trapped Bose gas has so far been described in terms of the grand 
canonical ensemble. This implies an open system exchanging energy and 
particles with a reservoir, where the average particle number is given by 
the chemical potential. 

The experiments are performed at fixed energy and particle number, and should 
therefore be described in terms of the microcanonical ensemble. The derivation
of the corresponding partition function is in general a difficult task, which 
involves complicated combinatorial problems. Using the canonical ensemble 
simplifies the calculations considerably, and still satisfies the constraint 
of fixed particle number. 

The different ensembles are known to give the same predictions for 
thermodynamic quantities in the large $N$ limit~\cite{Pathria}. 
Below we compare the results for the grand canonical and canonical ensembles. 
The results for the specific heat turn out to agree well. 
However, for the condensate fraction we find significant deviations.\\

Assuring that each configuration is counted only once, the partition function
in the canonical ensemble can be written as
\begin{eqnarray}
Z = \sum_{n_1=0}^{\infty}\cdots\sum_{n_N=n_{N-1}}^{\infty}
x^{n_{1}+n_{2}+...+n_{N}} 
= \prod_{i=1}^N\frac{1}{1 - x^i}.
\end{eqnarray}
Here, $x = \exp(-\beta\omega)$. The internal energy 
is now easily found to be
\begin{eqnarray}
U = -\frac{\partial}{\partial\beta}\ln Z = \sum_{i=1}^N\frac{i\omega x^i}{1 - x^i},
\end{eqnarray}
and the specific heat becomes
\begin{eqnarray}
C_V = \frac{\partial U}{\partial T} = 
\sum_{i=1}^N\frac{(i\beta\omega)^2x^i}{(1 - x^i)^2}.
\end{eqnarray}
The derivation of the condensate fraction is a little more involved: 
It can be written as
\begin{eqnarray}
N_0 = \sum_{k=1}^Nk p_k,
\end{eqnarray}
where $p_k$ is the probability of finding $k$ particles in the lowest energy 
level. This probability is found summing all contributing configurations:
\begin{eqnarray}
p_k = \frac{1}{Z}\sum_{n_{k+1}=1}^{\infty}\cdots\sum_{n_N=n_{N-1}}^{\infty}
x^{n_{k+1}+n_{k+2}+...+n_{N}} = x^{N-k}\prod_{i=0}^{k-1}(1 - x^{N-i}).
\end{eqnarray}

In Fig. 5 we have plotted the condensate fraction and the specific heat 
for $N = 10^3$ and $N = 10^6$, as given by the two ensembles. The difference 
in specific heat is hardly visible for the lower particle number, and for 
$N = 10^6$ the results of the two different ensembles agree to six decimal 
places. 
\begin{figure}[tbh] 
\begin{center}
\mbox{\psfig{figure=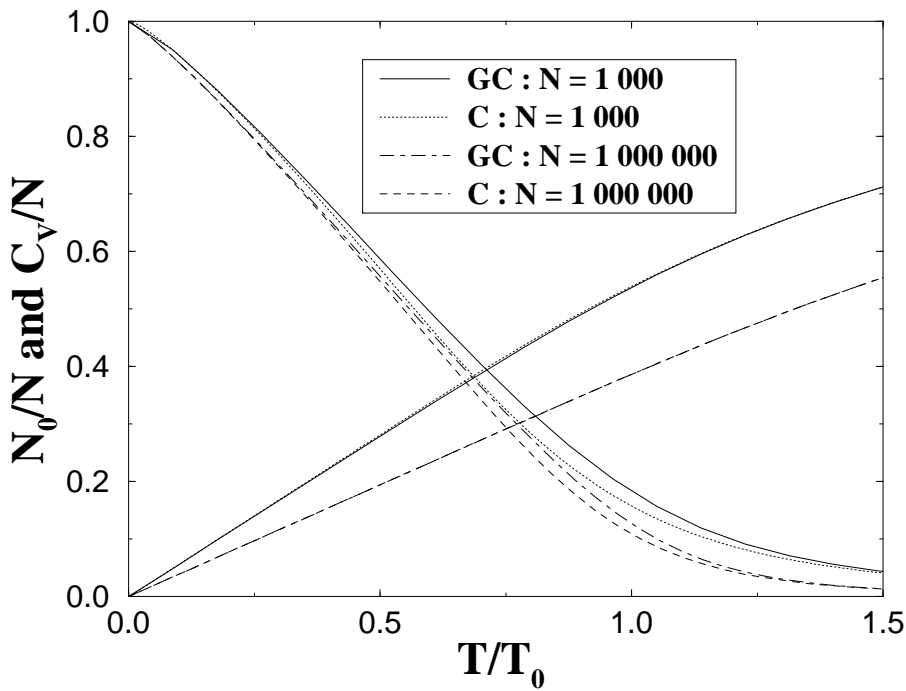,height=9cm,angle=270}}
\end{center}
\caption{\protect\footnotesize The condensate fraction and the specific heat in different ensembles
as a function of $T/T_{0}$ and for different values of $N$.}
\label{1d2}
\end{figure}

The situation for the condensate fraction is very different. One can show 
\cite{Haugset} that the error made calculating $N_0$ in the grand canonical 
ensemble is, at most, of the order ${\cal O}(1/\ln N)$. This is seen from the 
numerical results given in Fig. 5, where a significant deviation is found at intermediate temperatures even for $N=10^6$. 

This result is in contrast to the case of a free gas in three 
dimensions, where the deviation is known to go like $(\ln N)/N$ \cite{Hauge}. 
A more complete discussion of the finite $N$ effects for this system in the 
canonical ensemble will be given elsewhere \cite{Haugset}.

\subsection{Two Dimensions}
For completeness we also include results for an isotropic two-dimensional 
trapped gas. The equation for the particle number here takes the form
\begin{eqnarray}
N = N_0 + \frac{1}{b^2}\mbox{Li}_2(\bar{\lambda}) + \frac{2}{b}\mbox{Li}_1(\bar{\lambda}) 
+ \frac{11}{12}\mbox{Li}_0(\bar{\lambda}) + \frac{b}{6}\mbox{Li}_{-1}(\bar{\lambda}).
\end{eqnarray}
The expression for the internal energy reads
\begin{eqnarray}
U = \omega\left[\frac{2}{b^3}\mbox{Li}_3(\bar{\lambda}) + \frac{3}{b^2}\mbox{Li}_2(\bar{\lambda}) 
+ \frac{2}{b}\mbox{Li}_1(\bar{\lambda}) + \frac{3}{4}\mbox{Li}_0(\bar{\lambda}) 
+ \frac{b}{6}\mbox{Li}_{-1}(\bar{\lambda})\right].
\end{eqnarray}\\\\
For large $N$ one may define an approximate condensation temperature setting 
$N_0=0$ and $\lambda = 1$. The leading term at high temperature then 
gives~\cite{KvD}
\begin{eqnarray}
N = \left(\frac{T_0}{\omega}\right)^2Li_2(1) \hspace{1cm} \mbox{or} 
\hspace{1cm} \frac{T_0}{\omega} = \sqrt{\frac{N}{\zeta(2)}} 
= \frac{1}{\pi}\sqrt{6N}.
\end{eqnarray}
The limiting value at large $N$ for the condensate fraction is
\begin{eqnarray}
\frac{N_0}{N} = 1 - \left(\frac{T}{T_0}\right)^2.
\end{eqnarray}
At temperatures just below $T_0$ and large $N$ the energy takes the value 
$U = 2\omega\zeta(3)(T/\omega)^3$. This gives a specific heat 
\begin{eqnarray}
C_V(T\rightarrow T_0^-) = 6N\zeta(3)/\zeta(2).
\end{eqnarray}
Above $T_0$ the leading terms for $C_V$ are, again in the large $N$ limit
\begin{eqnarray}
\frac{C_V}{N} = 6\frac{Li_3(\lambda)}{Li_2(\lambda)} - 
4\frac{Li_2(\lambda)}{Li_1(\lambda)}.
\end{eqnarray}
Setting $\lambda=1$, we see that the second term vanishes, and the 
specific heat approaches the same value as found above. There is thus no 
jump in the specific heat even in the large $N$ limit in two dimensions.

In Fig. 6 we have plotted the condensate fraction  for different 
values of $N$. The transition sharpens as $N$ grows. The specific 
heat displayed in Fig. 7 is seen to develop a sharp, though continuous 
peak with increasing $N$.\\
\begin{figure}[tbh] 
\begin{center}
\mbox{\psfig{figure=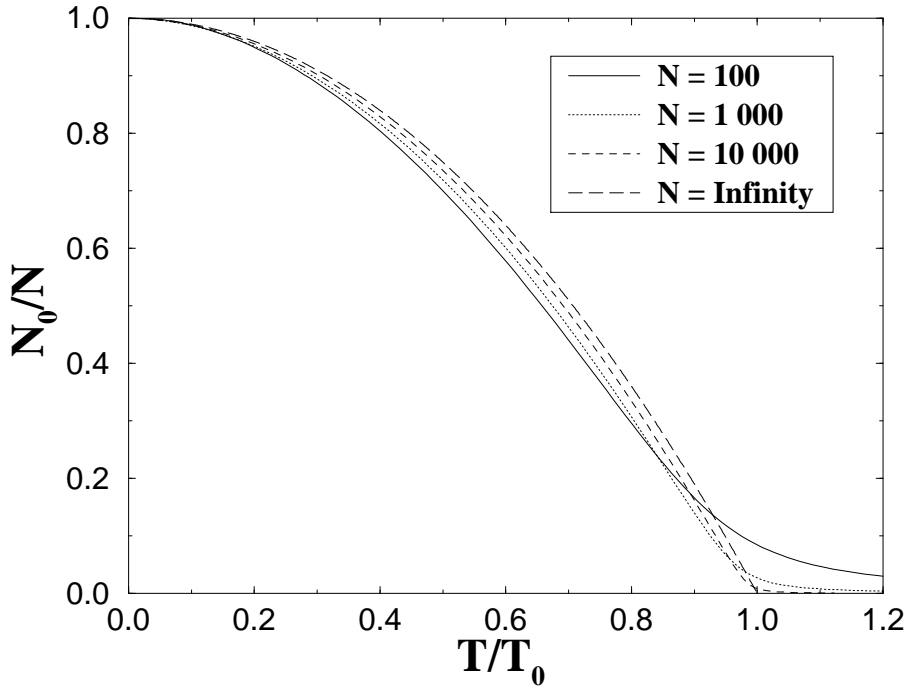,height=9cm,angle=270}}
\end{center}
\caption{\protect\footnotesize The condensate fraction for different values of the particle number 
$N$ as a function of $T/T_{0}$.}
\label{2d2}
\end{figure}
\begin{figure}[tbh] 
\begin{center}
\mbox{\psfig{figure=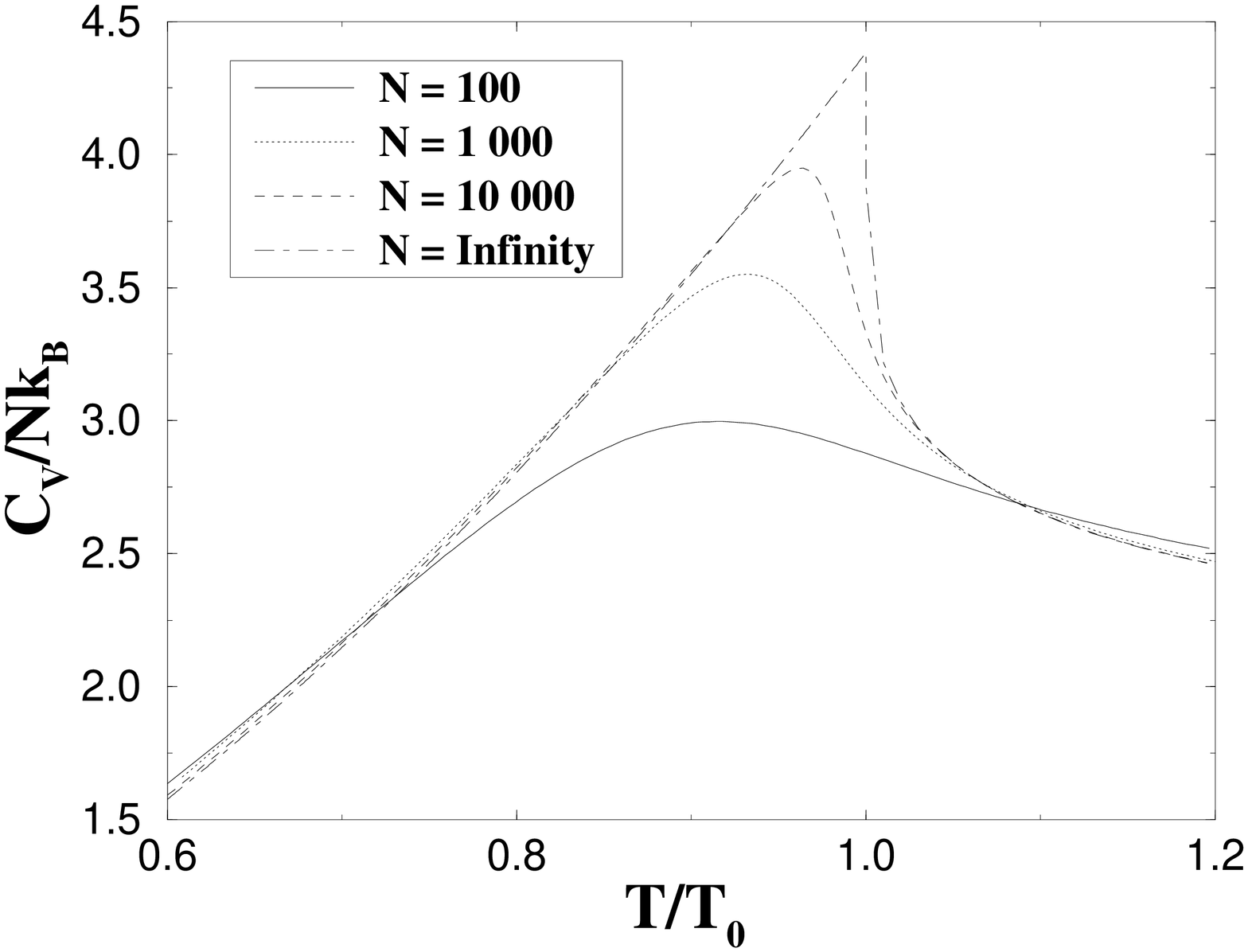,height=9cm,angle=270}}
\end{center}
\caption{\protect\footnotesize The specific heat for different values of the particle number 
$N$ as a function of $T/T_{0}$.}
\label{2d1}
\end{figure}

This system behaves similarly to the one in three dimensions: The specific 
heat has a maximum, and the condensate has a powerlike fall off 
in the large $N$ limit
($N_{0}\sim 1 - (T/T_0)^2$ in 2$D$ and $N_{0}\sim 1 - (T/T_0)^3$ in 3$D$).
Thus, as opposed to the free case, a 
two-dimensional trapped Bose gas undergoes a phase transition to a 
condensed state at {\it finite} temperature\footnote{During the preparation 
of this paper we came across a recent preprint by 
W. J. Mullin~\cite{Mullin}, where the same conclusions concerning BEC in lower dimensions 
are drawn.}.

\section{Conclusions}
In the present paper we have discussed the thermodynamics of an ideal Bose gas
trapped in an anisotropic harmonic well in some detail. We have investigated
the condensate fraction and the heat capacity for realistic values of the
frequencies by numerical as well as analytical methods. There is an
excellent agreement between the two methods in all dimensions.
Moreover, we have
studied the highly anisotropic trap where one of the frequencies is much
smaller than the others. This system should be realizable in future
experiments and it is expected to be effectively 
one-dimensional for low temperatures. We have therefore 
compared the condensate and the heat capacity for this system 
with the corresponding quantities in the one-dimensional trap. The overall
agreement for low temperatures confirms the above expectations.
We have also considered  thermodynamic quantities in different ensembles
at finite $N$. The most remarkable result is the deviation of the 
number of particles in the ground state.

Finally, we would like to mention several possible improvements of the present
treatment of BEC. The most obvious is the inclusion of interactions between
the atoms in the gas. This could either be based on perturbation theory
or some nonperturbative method. There are already a few articles on the
subject~\cite{string}, but more work needs to be done.
Secondly, one should use the microcanocial ensemble in order to compare 
with experiment. It is essentially a combinatorial challenge to do so.
More insight into finite $N$ effects of the ideal system
is also of interest, and a detailed
discussion will be presented in a follow up paper by one of the present 
authors~\cite{Haugset}.



\end{document}